\documentclass[twocolumn]{aastex631}
\shorttitle{Dynamics of Four Triple Systems}

\begin{document}

% for float placement:
\renewcommand{\topfraction}{1.0}
\renewcommand{\bottomfraction}{1.0}
\renewcommand{\textfraction}{0.0}

\newcommand{\kms}{km~s$^{-1}$\,}
\newcommand{\masyr}{mas~yr$^{-1}$\,}
\newcommand{\msun}{$M_\odot$\,}

\title{Dynamics of Four Triple Systems}

\author{Andrei Tokovinin}
\affiliation{Cerro Tololo Inter-American Observatory | NSF's NOIRLab
Casilla 603, La Serena, Chile}
\email{andrei.tokovinin@noirlab.edu}

\begin{abstract}
Orbital  motions in  four hierarchical  stellar systems  discovered by
speckle interferometry are studied.  Their inner orbits are relatively
well constrained, while  the long outer orbits are  less certain.  The
eccentric and  misaligned inner  orbits in the  early-type hierarchies
$\epsilon$~Cha  (B9V,  central star  of  the  5 Myr  old  association,
$P=6.4$\,yr, $e=0.73$), and I~385 (A0V,  $P \sim 300$ yr, $e \sim0.8$)
suggest past  dynamical interactions. Their nearly  equal masses could
be  explained by  a  dynamical  decay of  a  2+2 quadruple  progenitor
consisting of four similar stars. However,  there is no evidence of the
associated recoil,  so similar masses  could be just a  consequence of
accretion from  the same  core.  The  other two  hiearchies, HIP~32475
(F0IV, inner period 12.2 yr) and HIP~42910 (K7V, inner period 6.8 yr),
have smaller  masses and are double  twins where both inner  and outer
mass ratios are close to one.   A double twin could either result from
a merger of  one inner pair in a  2+2 quadruple or can be  formed by a
successive fragmentation followed by accretion.
\end{abstract}


%\maketitle

%---------------------------------------------------------
\section{Introduction}
\label{sec:intro}

Multiple stellar systems are very diverse, ranging from compact planar
worlds, where three  or four stars are tightly packed  within 1 au, to
wide  systems  of  0.1\,pc  scale,  often  found  in  non-hierarchical
configurations; see  \citet{Mult2021} for a review.   Hierarchies with
separations  of 1--100  au,  in the  middle of  this  range, are  more
typical.   Their  dynamics   (periods,  eccentricities,  mutual  orbit
orientation) bears imprints of the formation processes.  However, only
for a tiny fraction of known triple systems the inner and outer orbits
could  be  determined or  constrained  owing  to long  (centuries  and
millenia)  outer periods  and insufficient  data.  It  is increasingly
clear that hiearchies were formed via several different channels.

In  this   work,  orbits   are  determined   for  four   such  systems
(Table~\ref{tab:list}),   continuing  similar   studies  reported   in
\citep{Trip2021,TL20,Dancingtwins,TL2017}.    Inner  pairs   in  these
systems were  discovered a decade  ago by speckle  interferometry, and
the  data accumulated  to date  allow calculation  of the  first inner
orbits.   The outer  orbits are  not yet  fully covered.   Two systems
($\epsilon$~Cha and  I~385) have similar components  of early spectral
type arranged  in apparently non-hierarchical  configurations.  Their
inner  orbits have  large  eccentricities,  suggesting that  dynamical
interactions played  a major role.  The other two triples  contain
solar-type stars and are double twins  where a pair of  similar low-mass
stars orbits the primary component with mass comparable to the mass of
the pair.  Despite apparent similarity, the two double twins have very
different dynamics:  the first has quasi-circular  and aligned orbits,
while in the other the inner orbit is highly eccentric.

\begin{deluxetable*}{c l cc cc c c l  }
\tabletypesize{\scriptsize}
%\tablenum{7}
\tablewidth{0pt}
\tablecaption{List of Multiple Systems
\label{tab:list} }
\tablehead{
\colhead{WDS} &
\colhead{Name} &
\colhead{HIP} & 
\colhead{HD} & 
\colhead{$V$} &
\colhead{$\varpi$\tablenotemark{a} } & 
\colhead{$P_{\rm out}$} &
\colhead{$P_{\rm in}$ } &
\colhead{Masses} 
 \\
\colhead{(J2000)} &        &  & &
\colhead{(mag)} & 
\colhead{(mas)} & 
 \colhead{(yr)} &
 \colhead{(yr)} &
 \colhead{(\msun)}
}
\startdata
06467$+$0822 & HDS~940 A,BC    & 32475 & 49015    & 7.04 & 13.75 G  & 80.3    & 12.2 &  1.40+(0.69+0.65) \\
08447$-$2126 & HDS~1260 A,BC  & 42910 & \ldots & 10.19 & 27.51 G    & 125    & 6.9  &  0.72+(0.37+0.36) \\
11596$-$7813 & $\epsilon$~Cha & 58484 & 104174 & 4.90  & 9.02 H     & 750:    & 6.4  &  (2.57+2.45)+2.54 \\
17248$-$5913 & I~385 AD,B     & 85216 & 157081 & 7.25  & 3.85 G     & 2000:   & 300:  &  (2.32+2.03)+1.99 \\
\enddata
\tablenotetext{a}{Parallax codes: G --- Gaia DR3 \citep{Gaia3}, H --- Hipparcos \citep{HIP2}.}
\end{deluxetable*}

The   input    data   and   methods   are    briefly   introduced   in
Section~\ref{sec:methods}. Sections  \ref{sec:echa}--\ref{sec:HDS} are
devoted to individual systems. Their possible formation scenarios are
discussed in Section~\ref{sec:disc}.

%---------------------------------------------------------
\section{Data and methods}
\label{sec:methods}

%---------------------------------------------------------
\subsection{Speckle Interferometry}
\label{sec:speckle}

In the hierarchies studied here  inner subsystems have been discovered
by  speckle interferometry  with  the  high-resolution camera  (HRCam)
working on  the 4 m  telescopes SOAR (Southern  Astrophysical Research
Telescope) and Blanco  located in Chile. HRCam, in use  since 2007, is
based on  the electron multiplication CCD  detectors.  The instrument,
data processing,  and performance are covered  in \citep{TMH10,HRCam}.
The latest series of measurements and references to prior observations
can be  found in \citep{Tokovinin2022}.  Image cubes  of 200$\times$200
pixels and 400 frames are recorded  mostly in the $y$ (543/22\,nm) and
$I$ (824/170\,nm) filters  with an exposure time of  25\,ms or shorter
and   a  pixel   scale   of   15\,mas.   In   the   $y$  filter,   the
difffraction-limited resolution  of 30\,mas can be  attained, and even
closer separations can be measured via careful data modeling.  On the
other hand,  the $I$ filter  offers deeper magnitude limit  and better
sensitivity to faint, red companions.

\begin{figure}
\epsscale{1.1}
%\plotone{ACF-all.eps}
\plotone{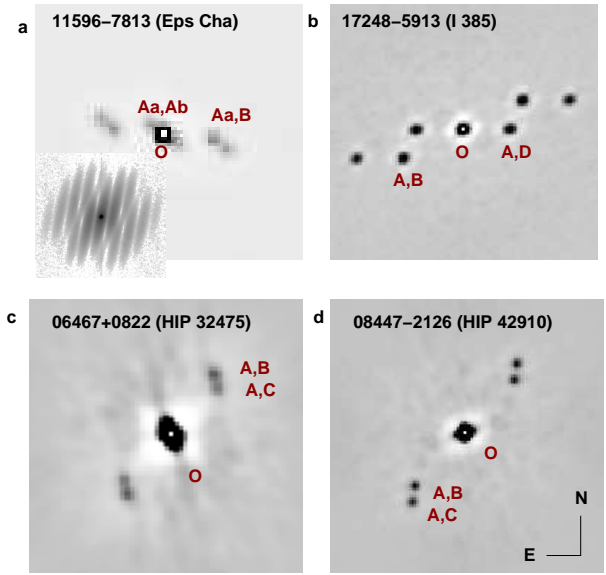}
\caption{Speckle  ACFs  of triple  stars  recorded  at SOAR  (negative
  intensity rendering, standard orientation, arbitrary scale). In each
  panel,  the white  dot O  marks  the center,  other labels  indicate
  correct peaks  corresponding to  components' pairs (other  peaks are
  symmetric    counterparts   or    complementary   pairs).     (a)
  $\epsilon$~Cha  on 2022.05,  separations 0\farcs054  and 0\farcs147;
  insert shows the  power spectrum. (b) I~385  on 2022.31, separations
  0\farcs29  and 0\farcs41.   (c) 06478+0822  on 2022.77,  separations
  0\farcs073 and  0\farcs34. (c) 08447$-$2126 on  2022.28, separations
  0\farcs067 and 0\farcs48.
\label{fig:ACF}  }
\end{figure}

Image  cubes are  processed by  the standard  speckle method  based on
calculation of  the spatial power spectrum  and image auto-correlation
function (ACF)  derived from the  latter.  The 180\degr  ~ambiguity of
position angles inherent to this  method is resolved by examination of
the  shift-and-add (``lucky'')  images  and by  comparison with  prior
data. In a  triple star, the angles of subsystems  are related, so the
better-defined   orientation  of   the  outer   pair  constrains   the
orientation of the  inner subsystem.  Figure~\ref{fig:ACF} illustrates
speckle  data on  the triple  systems studied  here.  Recall  that the
positions and relative photometry are determined by modeling the power
spectra, not by fitting the ACF peaks.

%---------------------------------------------------------
\subsection{Orbit Calculation}
\label{sec:orbit3}

As in the previous papers, an  IDL code that fits simultaneously inner
and   outer    orbits   in   a    triple   system   has    been   used
\citep{orbit3}.\footnote{Codebase:
  \url{http://dx.doi.org/10.5281/zenodo.321854}}    The   method    is
presented in \citet{TL2017}. No useful radial velocity measurements are
available   for  the   systems  studied   here,  so   only  positional
measurements are used.  The weights  are inversely proportional to the
squares  of adopted  measurement  errors which  range  from 2\,mas  to
0\farcs05   and    more   \citep[see][for   further    discussion   of
  weighting]{Trip2021}.

Motion in  a triple system  can be  described by two  Keplerian orbits
only approximately, but  the effects of mutual dynamics  are too small
to be detectable with the current  data.  The code fits 14 elements of
both  orbits and  the additional  parameter $f$  -- the  wobble factor,
ratio of the astrometric  wobble axis to the full axis of
the inner  orbit.  For resolved triples,  $f = q/(1+q)$, where  $q$ is
the  inner mass  ratio.  When  the inner  subsystem  is not  resolved,
measurements of the outer pair refer  to the photo-center of the inner
pair, and the wobble amplitude corresponds  to a smaller factor $f^* =
q/(1+q)  - r/(1+r)$,  where  $r$  is the  flux  ratio.   The code  {\tt
  orbit4.pro} can  accept a mixture  of resolved and  unresolved outer
positions; it adopts a fixed  ratio $f^*/f$, specified for each system
as input parameter.

For  the two  early-type  outer   pairs  discovered   visually,  position
measurements at SOAR  are complemented by the  historic micrometer and
speckle data retrieved from the  Washington Double Star (WDS) database
\citep{WDS} on my request. Although such data extend the time coverage
to  almost 200  yr (for  $\epsilon$~Cha), it  is still  too short  for
constraining outer periods of  several centuries.  To avoid the degeneracy
of outer  orbits, some  elements are fixed  to reasonable  values that
agree with the estimated masses.   The resulting outer orbits are only
representative;  however,  they are  still  useful  for the assessment  of
mutual dynamics.

The elements of inner and outer  orbits in the selected triple systems
are  given in  Table~\ref{tab:orb}  in standard  notation.  Given  the
uncertain nature of outer orbits,  the formal errors of their elements
are meaningless, so  they are not provided.   Individual positions and
their residuals to orbits are listed in Table~\ref{tab:obs}, available
in full  electronically.  Compared  to the  published HRCam  data, the
positions  are  corrected  for  the small  systematics  determined  in
\citep{Tokovinin2022} and,  in a  few cases, re-processed.  The second
column indicates the subsystem; for  example, A,BC refers to the angle
and separation between  A and unresolved pair BC, while  A,B refers to
the position of resolved component B relative to A.

\begin{deluxetable*}{c c r rrr rr l }
\tabletypesize{\scriptsize}
%\tablenum{7}
\tablewidth{0pt}
\tablecaption{Positional Measurements and Residuals \label{tab:obs}}
\tablehead{
\colhead{WDS} & 
\colhead{System} & 
\colhead{$T$} &
\colhead{$\theta$} & 
\colhead{$\rho$} &
\colhead{$\sigma$} & 
\colhead{O$-$C$_\theta$} & 
\colhead{O$-$C$_\rho$} &
\colhead{Ref.\tablenotemark{a}} \\
& & 
\colhead{(yr)} & 
\colhead{(\degr)} &
\colhead{(\arcsec)} & 
\colhead{(\arcsec)} & 
\colhead{(\degr)} &
\colhead{(\arcsec)} &
}
\startdata
06467+0822 & B,C &  2015.9063  &   2.6 & 0.0823 &  0.005  &  -1.0 & 0.0037 & S \\
06467+0822 & B,C &  2015.9063  &   1.5 & 0.0835 &  0.005  &  -2.1 & 0.0049 & S \\
06467+0822 & B,C &  2016.9575  &  33.5 & 0.0812 &  0.005  &  3.8 & -0.0057 & S \\
\enddata
\tablenotetext{a}{
H: Hipparcos;
M: visual micrometer measurement;
S: speckle interferometry at SOAR;
s: speckle interferometry at other telescopes.}
\tablenotetext{}{(This table is available in its entirety in
  machine-readable form) }
\end{deluxetable*}

\begin{deluxetable*}{ l  c rrr rrr r r  }
\tabletypesize{\scriptsize}
%\tablenum{7}
\tablewidth{0pt}
\tablecaption{Orbital Elements \label{tab:orb}}
\tablehead{
%\multicolumn{3}{c}{Discoverer} &
\colhead{WDS} &
\colhead{System} &
\colhead{$P$} & 
\colhead{$T  $} &
\colhead{$e$} & 
\colhead{$a$} & 
\colhead{$\Omega$} &
\colhead{$\omega$} &
\colhead{$i$}  &
\colhead{$f$} 
 \\
 & &   
\colhead{(yr)} & 
\colhead{(yr)} &
\colhead{ } & 
\colhead{($''$)} & 
\colhead{(\degr)} &
\colhead{(\degr)} &
\colhead{(\degr)} & 
}
\startdata
06467+0822  & B,C  &  12.20   & 2011.70  & 0.095     & 0.0827     &  54.7  & 171.8     & 32.3    & $-$0.50 \\
             &   & $\pm$0.38  &$\pm$0.06 &$\pm$0.032 &$\pm$0.0033 &$\pm$8.0 &$\pm$20.5 &$\pm$5.3 & $\pm$0.03 \\
06467+0822  & A,BC &  80.3   & 2029.0   & 0.072     & 0.382     &  23.6   & 324.7    & 30.7    & \ldots \\
             &    & $\pm$1.9  &$\pm$5.7 &$\pm$0.055 &$\pm$0.108 &$\pm$7.5 &$\pm$20.0 &$\pm$5.7 & \ldots \\
08447$-$2126 & B,C &  6.845  & 2016.416    & 0.948      & 0.0889    &  8.6     &  0 & 160.0 & $-$0.48 \\
          &   & $\pm$0.034   &$\pm$0.023 & $\pm$0.006 &$\pm$0.009 & $\pm$0.6 & fixed & fixed & $\pm$0.03  \\
08447$-$2126 & A,BC  &  125  & 2020.05  & 0.292    & 0.783 &  26.3 & 235.0   & 160.0 & \ldots \\
11596$-$7813 & Aa,Ab  & 6.43  & 2018.57  & 0.733   & 0.0541 &  21.1 & 116.2        & 111.2 & 0.48  \\
             &   & $\pm$0.09  &$\pm$0.06&$\pm$0.020&$\pm$0.0022&$\pm$1.5 &$\pm$2.2&$\pm$1.8 & $\pm$0.02 \\
11596$-$7813 & A,B     & 751  & 1837.4  & 0.0    & 1.481  &  181.4 &  0       & 83.5 & \ldots  \\
11596$-$7813 & A,B     & 460  & 2061.3  & 0.75   & 1.092 &  173.8 & 213.6    & 78.2 & \ldots  \\
%             &   & fixed      &$\pm$4.8 & fixed  &$\pm$0.081&$\pm$0.9 &$\pm$4.2&$\pm$1.8   \\
17248$-$5913 & A,D & 300      &  1969.0  & 0.80  & 0.280  & 271.5  & 242.0   & 90.0 & 0.43 \\
%             &
17248$-$5913 & AD,B &2000    & 1566     & 0.12   & 1.13   & 243.6  & 12.1  & 111.0 & \ldots \\
%             &
\enddata
\end{deluxetable*}

%---------------------------------------------------------
\section{Epsilon Chamaeleontis}
\label{sec:echa}

The bright ($V=4.90$, $K=4.98$  mag) B9V star $\epsilon$~Cha (HR~4583,
HD 104174,  HIP~58484, WDS J11596$-$7813)  is the central star  of the
young \citep[5$\pm$2  Myr,][]{Dickson2021} $\epsilon$  Cha association
located at $\sim$100 pc  average distance \citep{Murphy2013}. The star
has been  resolved in 1835.93  into a 1\farcs6 binary  with comparably
bright components  by \citet{Herschel1847} and designated  as HJ~4486.
Subsequent  monitoring with visual micrometers  revealed a slowly
decreasing   separation  with   little  change   in  position   angle.
Speckle-interferometric  and  Hipparcos   measurements  in  the  1990s
documented a separation of $\sim$0\farcs4.

\begin{figure*}
\epsscale{0.7}
%\plotone{Orbit.eps}
\plotone{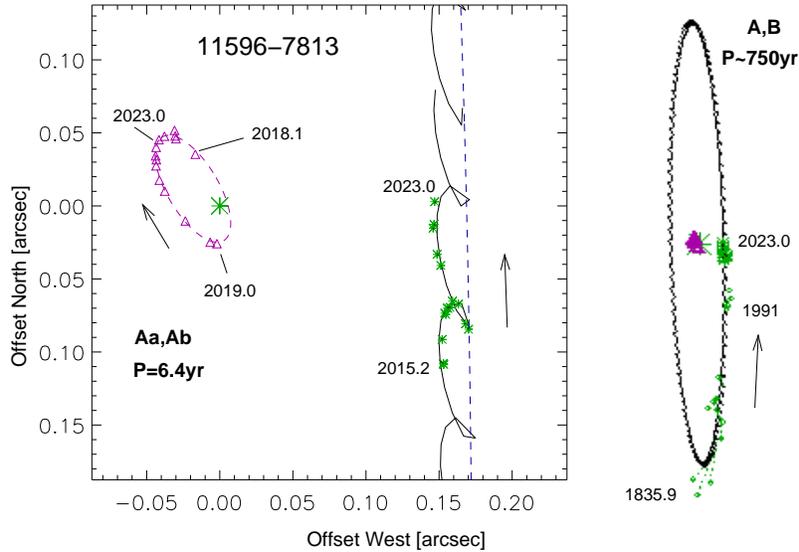}
\caption{Orbits of $\epsilon$~Cha.  The right-hand plot shows the full
  circular outer  orbit (crosses  denote the less  accurate micrometer
  measurements,  squares   show  the  resolved  speckle   data).   The
  left-hand plot  shows SOAR measurements  of the inner  pair (magenta
  ellipse and triangles)  and the wavy line  of the Aa,B motion  with the
  superimposed wobble. The blue dashed line shows outer orbit without
  wobble. 
\label{fig:echa-orb}
}
\end{figure*}

The pair  A,B was closing down  and lacked recent measurements,  so it
was  observed   at  SOAR   in  2015.25  on   request  by   Ross  Gould
\citep{Tok2016}.  Quite unexpectedly, $\epsilon$~Cha was revealed as a
tight triple consisting of similar stars (Figure~\ref{fig:ACF}a).  The
inner pair Aa,Ab  with a separation of 51\,mas was  expected to have a
short orbital period and, indeed, its fast orbital motion was detected
in  the  following  years  \citep{Briceno2017}.   In  2022  Aa,Ab  has
completed one full revolution since  its discovery, and its orbit with
a period of 6.4\,yr is determined here.

The fluxes of the three components  of $\epsilon$~Cha are similar, but not
exactly equal, which helps to  establish the orientation.  As shown in
Figure~\ref{fig:ACF}a, the ACF peak below B is slightly weaker than the
peak of B  itself, thus defining the orientation of  Ab relative to Aa
as indicated.  The  13 SOAR measurements in the $y$  filter average to
$\Delta y_{\rm  Aa,Ab} = 0.25$ mag  and have an rms scatter  of 0.04 mag.
At the same time,  $\Delta y_{\rm Aa,B} = 0.11$ mag  with a scatter of
0.07 mag.  The combined magnitude  of $V=4.90$ mag leads to the individual
$V$  magnitudes  of  Aa,  Ab,  and   B:  5.98,  6.23,  and  6.09  mag,
respectively.

Speckle interferometry  allows  a position  angle change  by
180\degr (flip), but only simultaneous flips of both pairs in a triple are
allowed.   The  orientation   of  A,B  is  defined   by  the  historic
measurements, thus fixing  the angle of Aa,Ab.  However,  when in 2019
Aa,Ab  closed  down  below  the   diffraction  limit,  the  ACF  peaks
overlapped  and it  was no  longer possible  to discriminate  reliably
between opposite  angles of  the inner  pair.  The  data of  2019 were
originally processed  under the assumption  that Ab is located  to the
north of Aa, extrapolating its  retrograde motion from the previous years.
However,  a  negative  $\Delta  y_{\rm  Aa,Ab}$  indicated  that  this
assumption was  incorrect, as also  confirmed by the orbit.   The SOAR
observations in 2019  were re-fitted with the  reversed orientation of
Aa,Ab, which also affected the measured positions of Aa,B.

The orbits  of Aa,Ab and A,B  were fitted jointly. Apart  from the WDS
data,  one speckle  measurement made  at  Gemini-S in  2017.4 is  used
\citep{Horch2019},  the rest  are  SOAR  measurements.  The  resulting
wobble factor $f = 0.48 \pm 0.02$  indicates that the masses of Aa and
Ab are equal,  $q_{\rm Aa,Ab} = 0.92 \pm 0.08$.   The first attempt to
compute the orbit  of Aa,Ab using wrong quadrants in  2019 resulted in
an unrealistically small  mass sum, but after  quadrant correction the
orbit  of  Aa,Ab  becomes   almost  perfect  (weighted  rms  residuals
0.9\,mas) and corresponds  to the inner mass sum  of 5.2$\pm$0.6 \msun
using the Hipparcos parallax of  9.02\,mas.  The outer orbit, however,
is not yet  constrained by the observed arc, allowing  a wide range of
solutions.  Two  orbits of  A,B are  listed in  Table~\ref{tab:orb}: a
circular one with  $P=751$~yr and an eccentric  orbit with $P=460$~yr.
The circular orbit is adopted below; it corresponds to the mass sum of
7.9  \msun.   Dynamical  stability  of  the   triple  system  requires
a separation  of  $>$0\farcs2 at  the  outer  periastron, so  the  outer
eccentricity should not exceed 0.8.

Note   that  the   inner  pair   moves  clockwise,   the  outer   pair
counterclockwise,  so the  two  orbits cannot  be coplanar.   However,
without identification of the correct  ascending nodes of both orbits,
the  mutual inclination  can  take two  possible  values, 156\degr  ~or
34\degr ~for the  circular outer orbit (156\degr ~and  41\degr ~for the
eccentric  one).   The  first  value  corresponds  to  counter-aligned
orbital  angular  momenta, while  the  second  implies only  a  modest
inclination.   It   is  likely  that  mutual   inclination  and  inner
eccentricity vary in Lidov-Kozai cycles. The large inner (and possibly
outer)  eccentricities attest,  indirectly,  to dynamical  interaction
between the orbits.

\begin{figure}
\epsscale{1.1}
%\plotone{Sky2.eps}
\plotone{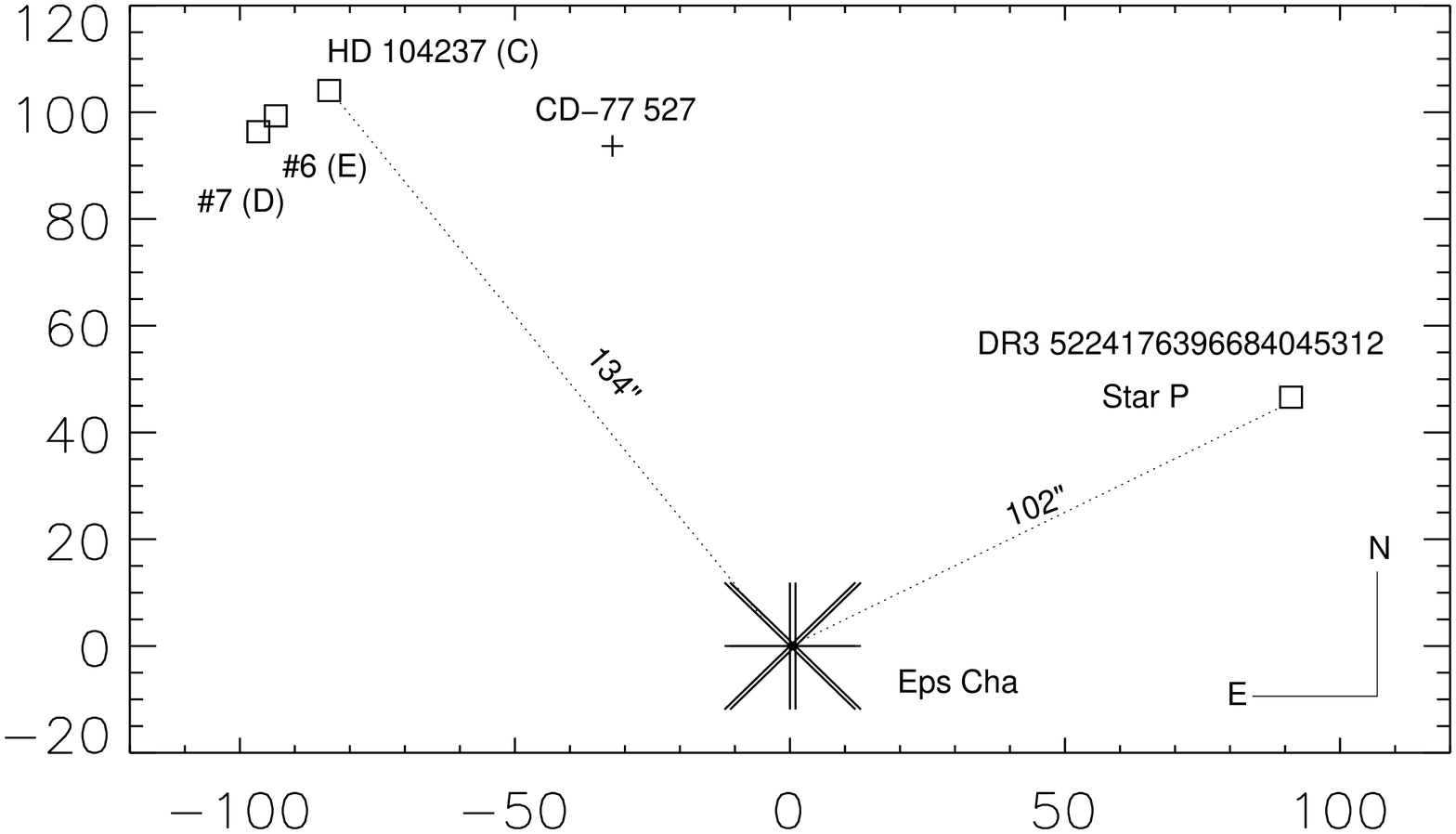}
\caption{Closest   neighbors   of   $\epsilon$~Cha  (axis   scale   in
  arcseconds) in Gaia DR3.
\label{fig:echa-sky}
}
\end{figure}

The  Multiple   Star  Catalog   \citep{MSC}  and  the   WDS  associate
$\epsilon$~Cha  with another  multiple system,  HD~104237 (HIP  58520,
DX~Cha,  $V=6.60$  mag,  A7Ve)  located  at  an  angular  distance  of
134\arcsec ~(projected  separation 15  kau or 0.07\,pc;  FGL~1 AB,C).
The projected separation implies an orbital period of $\sim$0.5 Myr if
these stars  are gravitationally  bound. HD~104237 is  a spectroscopic
binary with a period of 19.86  days that has been extensively studied;
it  is   accreting  from  a  circumbinary   disk  \citep{Dunhill2015}.
Furthermore, HD~104237 is surrounded by a swarm of five faint low-mass
stars within  15\arcsec ~according to \citet{Grady2004}  and Gaia; the
WDS code of this system is J12001$-$7812.  I looked for objects within
3\arcmin ~radius of $\epsilon$~Cha in Gaia DR3 \citep{Gaia3} and found
another  association member,  DR3 5224176396684045312  ($G=15.23$ mag,
parallax  9.413$\pm$0.025 mas,  proper motion $(-38.95,  -5.48)$
\masyr) at a distance of  101\arcsec, denoted provisionally as star P.
Location   of   the  neighbors   on   the   sky  is   illustrated   in
Figure~\ref{fig:echa-sky}.   The   star  CD$-$77~527   situated  between
$\epsilon$~Cha  and  HD~104237  does  not belong  to  the  association
(parallax 3.44 mas).

Gaia  does not  provide  parallax of  $\epsilon$~Cha, while  Hipparcos
measured 9.02$\pm$0.36 mas \citep[new reduction,][8.95$\pm$0.58 mas in
  the original  catalog]{HIP2}.  Comparison of the  Gaia and Hipparcos
positions gives the best estimate of the proper motion (PM), $(-42.85,
-10.14)$ \masyr.   Orbital motion of B  relative to A with  a speed of
12.8 \masyr  is directed  to the north.   However, for  multiples with
equal components the centers of mass and light coincide ($f^* =0$), so
correction of the  PM for the orbital motion is  not needed.  Gaia DR3
measured a parallax of  9.38$\pm$0.05 for HD~104237, compatible within
errors  with the  Hipparcos  parallax of  $\epsilon$~Cha.  The  formal
errors  of  Gaia and  Hipparcos  parallaxes  cannot be  fully  trusted
because  astrometry of  unresolved multiple  systems is  often biased.
Table~\ref{tab:neighbors} lists  parallaxes and  PMs of  the neighbors
found  in  Gaia.   Capital   letters  correspond  to  the  components'
designations in  the WDS and MSC.   The last column gives  the Reduced
Unit  Weight Error  (RUWE)  as  an indicator  of  the Gaia  astrometric
quality and potential subsystems.   The closest satellite of HD~104237
at 1\farcs4 separation (GRY~1 AF) has  no parallax and PM in Gaia DR3,
but the stability of its relative position over time proves that it is
bound.

The  projected  separations of  $\epsilon$~Cha  to  its neighbors  are
within 15\,kau, typical for wide binaries and triples and suggesting that
they may be bound.  However, the PM differences of $\sim$5 \masyr (2.5
\kms)  in Table~\ref{tab:neighbors}  appear highly  significant.  Note
also that two satellites of  HD~104237, $\epsilon$~Cha \#6 (E) and \#7
(D)  at  10\arcsec  ~and 15\arcsec  ~separations,  respectively,  have
measurably  different parallaxes,  implying  that this  pair might  be
$\sim$7\,pc closer to the Sun  and simply projects onto HD~104237. So,
the status of the neighbors remains undetermined.  They could be either
just independent members of the association or members of a bound (but
likely dynamically unstable) stellar system.

\begin{deluxetable}{l c r c c c c }
\tabletypesize{\scriptsize}
%\tablenum{7}
\tablewidth{0pt}
\tablecaption{Neighbors of $\epsilon$~Cha in Gaia DR3
\label{tab:neighbors} }
\tablehead{
\colhead{Name} &
\colhead{Sep.} &
\colhead{$G$} &
\colhead{$\varpi$} &
\colhead{$\mu^*_\alpha$} &
\colhead{$\mu_\delta$} &
\colhead{RUWE} \\
&
\colhead{(arcsec)} &
\colhead{(mag)} &
\colhead{(mas)} &
\multicolumn{2}{c}{(\masyr)} &
}
\startdata
$\epsilon$ Cha (AB) & 0     & 4.78  & 9.02: & $-$42.9  & $-$10.1 & \ldots \\ 
Star P          & 101.9 & 15.23 & 9.413 & $-$39.0  & $-$5.5 & 1.1 \\
HD 104237 (C)   & 133.9 &  6.56 & 9.380 & $-$39.3  & $-$5.8 & 2.1 \\
Eps Cha\#6 (E)  & 136.7 & 13.01 & 9.769 & $-$38.7  & $-$3.1 & 1.4 \\
Eps Cha\#7 (D)  & 136.8 & 12.78 & 9.980 & $-$42.8  & $-$4.2 & 1.6 
\enddata
\end{deluxetable}

\begin{figure}
%\plotone{CMD.eps}
\plotone{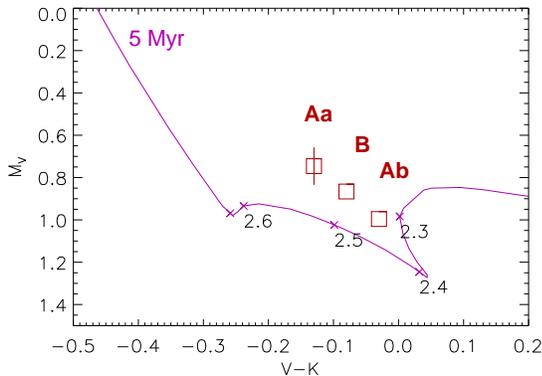}
\caption{Location  of $\epsilon$~Cha  components  Aa, B,  and Ab  (red
  squares) on  the $(V, V-K)$  CMD.  The  error bar  indicates the
  distance modulus uncertainty of $\pm$0.08 mag.  The magenta curve is
  a 5 Myr  PARSEC isochrone for solar  metallicity \citep{PARSEC} with
  masses marked by asterisks and numbers.
\label{fig:echa-cmd}
}
\end{figure}

The  relative photometry  of the  $\epsilon$~Cha components  allows to
place them on the color-magnitude diagram (CMD). The individual colors
are not measured, but, given  similar magnitudes, they should be close
to   the    combined   color    $V   -   K    =   -0.08$    mag.    In
Figure~\ref{fig:echa-cmd}, the colors are arbitrarily offset from this
value for illustration.  Overall, the Hipparcos distance, inner orbit,
and isochrone  lead to consistent  masses around 2.5  \msun.  However,
the isochrone  is not  monotonous in this  region, which  is sometimes
called  H-peak  \citep{Guo2021} and  corresponds  to  the ignition  of
hydrogen burning in young stars. At  5\,Myr age, the H-peak is located
at $M_G$  between 0  and 1  mag, matching  the absolute  magnitudes of
$\epsilon$~Cha components.   Given the  uncertainties in  the distance
and color,  potential inaccuracy of  the isochrone, and  its particular
shape, it is hazardous to infer masses from the isochrone; the masses
listed in Table~\ref{tab:list} are tentative.

Considering  the  young  age   of  $\epsilon$~Cha  and  the  continued
accretion on  its neighbor HD~104237,  it was worth checking  for the
presence of  hydrogen emissions  in the  spectrum. An  optical echelle
spectrum of $\epsilon$~Cha has been taken on 2022 February 25 with the
CHIRON echelle  spectrometer on  the 1.5-m  telescope at  Cerro Tololo
\citep{CHIRON}. The wide and deep  hydrogen Balmer lines have no signs
of emission, as established previously by \citet{Lyo2008}.  Apart from
that,  the  spectrum is  almost  featureless.   One notes  only  sharp
telluric absorptions in  the red part and a few  very shallow and wide
stellar lines.  Thus, any residual  gas around $\epsilon$~Cha has been
expelled and this  system is not accreting at  present.  Its potential
formation scenario is discussed below in Section~\ref{sec:disc}.

%---------------------------------------------------------
\section{Innes 385}
\label{sec:I385}

This  remarkable  quadruple  stellar  system is  known  as  HIP~85216,
HD~157081,  WDS J17248$-$5913,  and I~385.   The bright  visual triple
consisting  of the  0\farcs5 pair  A,B with  companion C  at 17\arcsec
~separation has been discovered by R.~Innes in 1901 \citep{Innes1905};
star C  has similar PM and  parallax, hence it belongs  to the system.
Another star E listed in the WDS (28\farcs01, 109\fdg3, $G=13.50$ mag)
is optical, as evidenced by its  distinct Gaia parallax (0.47 mas) and
PM  of $(-0.4,  -6.3)$  \masyr.   The inner  companion  D, similar  in
brightness  to A  and  B, has  been discovered  in  2008.5 by  speckle
interferometry (WSI~87 AD)  at 0\farcs26 separation, while  A,B was at
0\farcs39,  in a  spectacular  triangular configuration  \citep{TMH10}
shown in  Figure~\ref{fig:ACF}b.  The object  was regularly visited
by  the SOAR  speckle camera  since  its discovery.   During 14  years
(2008.5 -- 2022.3) the A,D pair has opened up slightly (from 0\farcs26
to 0\farcs28)  at a  rate of  2 \masyr  with constant  position angle,
while A,B  moved faster.   A preliminary analysis  of this  system was
provided in \cite{Tok2016}.

Examination of all available data has  led to the firm conclusion that
the  speckle-interferometric observation  of  this star  by the  CHARA
group in  1990.35 \citep{Hartkopf1993}  actually resolved  the triple.
Brian Mason consulted the archive and, indeed, the system was noted as
having ``possible third component''.  The position of A,D was measured
at 270\fdg2  and 0\farcs1994.  This pre-discovery  observation has not
been published  at the time, awaiting  for a confirmation; it  is used
here. Curiously, the CHARA team also  observed $\epsilon$~Cha at a 4 m
telescope in the  1990s three times, but they have  not discovered the
subsystem Aa,Ab.

Hipparcos   measured  the   parallax   of  A   as  3.15$\pm$0.96   mas
\citep{HIP2}.  Gaia  does not give parallax  of A because it  is not a
point  source.  However,  the accurate  Gaia  DR3 parallax  of star  C
(3.848$\pm$0.013\,mas) fixes  the distance  to this system.   The Gaia
astrometry  of  C is  of  good  quality  (RUWE=0.98).   The PM  of  C,
$(-8.989, -11.557)$  \masyr, matches  the Hipparcos  PM of  A, $(-8.7,
-14.5)$ \masyr; however, the  latter is a blend of A,  B, and D biased
by motion in the inner triple.  The PM of A derived from its Hipparcos
and Gaia positions is $(-5.39, -12.26)$ \masyr.

The median magnitude differences of A with B and D in the $y$ band are
0.49  and  0.42  mag,  respectively   (D  is  slightly  brighter  than
B). Considering  the combined  magnitude $V=7.25$ mag,  the individual
$V$ magnitudes of A, B, and  D are 8.16, 8.65, 8.58 mag, respectively.
The  absolute magnitudes  match  main-sequence stars  of masses  2.32,
1.99, and 2.03  \msun, and the combined spectral  type A0V corresponds
to a star of 2.3 \msun.

The fact that  the inner, closer pair A,D moves  slower than the wider
pair A,B is unusual.  \citet{Tok2016} proposed two explanations.  Star
D could  move on  a wide orbit  around A,B and  project onto  it. This
configuration has a low probability  and, moreover, the wide A,D orbit
could be dynamically  unstable with respect to the  outer companion C.
The other explanation  of apparently slow A,D motion is  because it is
near apastron  of an eccentric  and highly inclined orbit.   This more
natural  hypothesis  is  adopted   and  further  explored  here.   The
observations do  not cover the long  orbital periods of AD,B  and A,D;
the  short  observed segments  can  match  a  wide range  of  possible
orbits. The  question is  whether some of  those potential  orbits are
compatible with the distance and  estimated masses. To answer it, just
a pair of plausible orbits suffice.

\begin{figure}
\epsscale{1.1}
%\plotone{Orbit385.eps}
\plotone{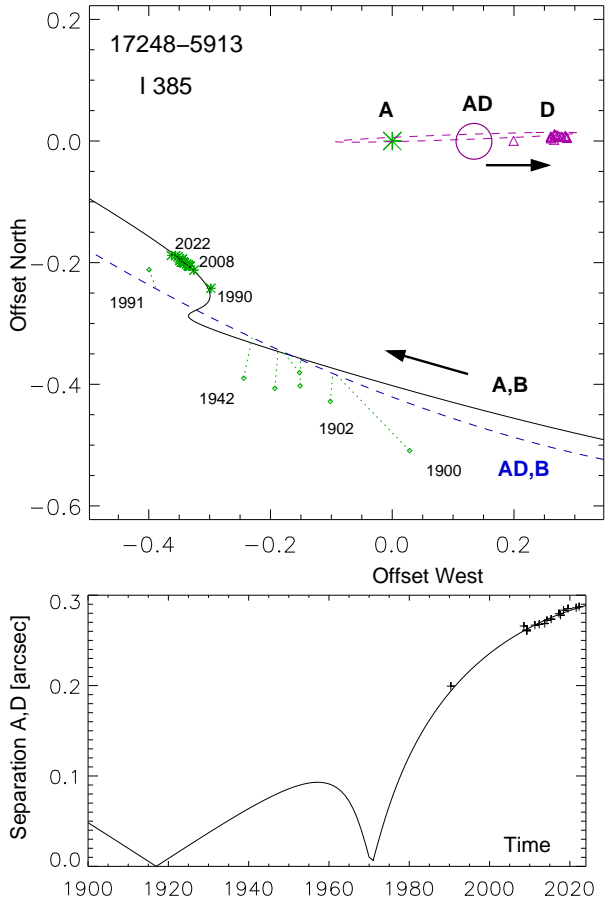}
\caption{Orbital motion of the inner  triple I~385. Top: motion of B
  relative to A (full line and  green asterisks) or relative to the AD
  photo-center  (blue dashed  line and  small diamonds).   The magenta
  line and triangles show  the orbit of A,D on the  same scale. Star A
  is  placed  at the  coordinate  origin.  The  plot on the bottom  shows
  angular separations  of A,D  vs.  time.  The  inner pair  was closer
  than  0\farcs1 throughout  most of  the  20th century  and for  this
  reason it has been missed  by  visual observers.   
\label{fig:I385-orb}
}
\end{figure}

First, I studied the motions of AD,B and A,D separately. A crude orbit
of AD,B with  $P=1244$ yr was suggested in  \citep{Tok2016}.  I assume
that  the  historic  micrometer  measurements of  AD,B  refer  to  the
unresolved inner  pair AD.  The  resolved speckle measurements  of A,B
and A,D were transformed by replacing  A with the average positions of
A and D (center of mass), assuming  that A and D are equal.  After the
initial fit,  the elements $P$ and  $a$ were fixed to  the values that
match the expected  mass sum of 6.3 \msun.  The  eccentricity of AD,B,
essentially  unconstrained,  is  fixed  to  a  small  value  (a  large
eccentricity would  render the inner pair  dynamically unstable).  The
actual values of $P,a$ can  be substantially larger than those adopted
here.

For the inner orbit of A,D,  I adopted the period of 300\,yr estimated
from  the projected  separation,  fixed $e=0.8$  and $i=90\degr$,  and
selected the  element $\omega$ to  obtain the  target mass sum  of 4.3
\msun.   The resulting  orbit fits  well the  observed slow  motion of
A,D. At present, the rate of  its opening up decreases, and in several
decades the pair will start to  close down.  In the final iteration, I
used the  {\tt orbit4.pro}  code to  model both  orbits simultaneously
(see Table~\ref{tab:orb}).  The masses  quoted above correspond to the
wobble factor $f=0.47$, while the fitted value is 0.43$\pm$0.13.  With
the parallax of  3.85\,mas, the inner and outer mass  sums are 4.4 and
6.3 \msun and, by design, match the photometric mass sums.

The  orbits are  illustrated in  Figure~\ref{fig:I385-orb}. One  notes
that the  first measurement  of A,B  by Innes  in 1900  is inaccurate.
Five micrometer  measurements of  AD,B in  1963--1979 are  omitted, as
well  as  the  highly  discrepant   measurement  by  Innes  in  1909.6
(discrepant   micrometer  measurements   are  common).    The  speckle
measurement   in  1990.35   by   \citet{Hartkopf1993}  at   0\farcs384
separation  matches the  resolved  position of  A,B  rather than  AD,B
(indeed, the triple was resolved at the time but not announced), while
the  Hipparcos  position in  1991.25  at  0\farcs452 better  fits  the
unresolved pair AD,B; it was likely biased by the triple nature of the
source.  The tentative orbits demonstrate  that the slow motion of A,D
is  compatible  with an  edge-on  eccentric  orbit.  This  orbit  also
explains why  the triple has  not been discovered  earlier: throughout
most  of  the  20th  century  A,D remained  too  close  for  a  visual
resolution.

%---------------------------------------------------------
\section{Double Twins HIP 32475 and HIP 42910}
\label{sec:HDS}

\begin{figure}
\epsscale{1.0}
%\plotone{Orb32475.eps}
\plotone{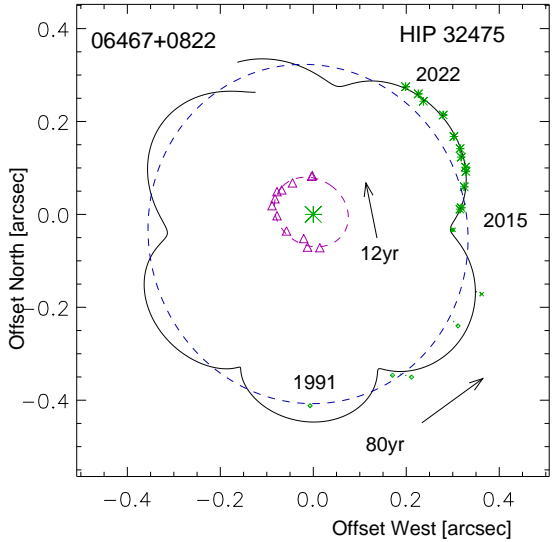}
\caption{Orbits of HIP 32475 with periods  of 12.2 yr and 80.3 yr. The
  blue dashed line marks the outer orbit without wobble that describes
  motion of  BC around  A, the  black solid  line is  the motion  of B
  relative to A.  The orbit of B,C is plotted around coordinate origin
  on the same scale by the magenta line and triangles.
\label{fig:32475-orb}
}
\end{figure}

\begin{figure}
\epsscale{1.0}
%\plotone{Orb42910.eps}
\plotone{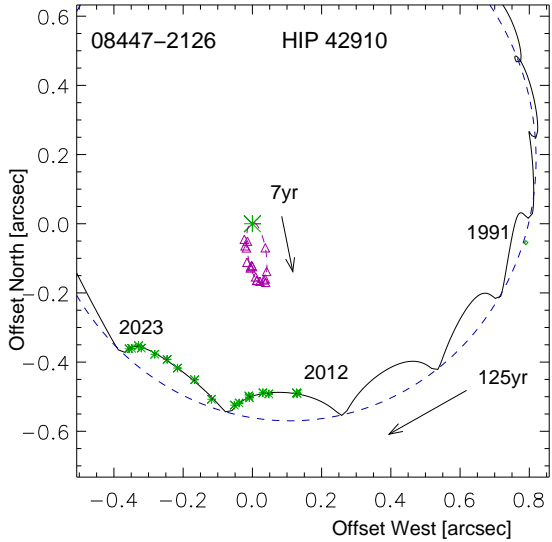}
\caption{Orbits of HIP 42910; periods 6.8\,yr  and 125\,yr.
\label{fig:42910-orb}
}
\end{figure}

The  two triple  systems featured  in  this Section  have some  common
features. Both are double twins where a more massive primary star A is
orbited  by a  twin secondary  pair of  low-mass stars  B and  C.  The
magnitude difference of B and C  relative to A is substantial, about 3
mag,  as  in other  similar  double  twins \citep{Dancingtwins}.   Yet
another similarity are moderate ratios of the outer and inner periods.

The outer pairs in these two  systems were discovered by Hipparcos and
are  named  HDS~940 and  HDS~1260,  respectively,  in the  WDS.   Both
secondaries were resolved  into close pairs at SOAR in  2014 and 2015.
Independently,  \citet{Horch2017}  discovered  the  triple  nature  of
HIP~42910 in  2012.  This team  also observed  HIP~32475 at the  3.5 m
WIYN  telescope five  times from  1998  to 2012.  They published  only
measurements  of  the  outer  pair  A,BC  and  apparently  missed  the
subsystem.  Figure~\ref{fig:ACF}  shows typical speckle ACFs  of these
triples  in the  $I$ band.   Both systems  are not  resolved by  Gaia.
Orbital motion  causes an increased astrometric  noise and potentially
affects parallaxes, although the bias caused by the century-long outer
orbits might be small.

The   orbits   are    plotted   in   Figures~\ref{fig:32475-orb}   and
\ref{fig:42910-orb}    and    their     elements    are    given    in
Table~\ref{tab:orb}.  The positional  measurements come from Hipparcos
(outer   pairs,  epoch   1991.25),  publications   by  Horch   et  al.
\citep[e.g.][]{Horch2017}, and SOAR. The coverage of both inner orbits
is  adequate, but  the outer  arcs  are covered  only partially.   The
shorter 80 yr outer orbit of HIP~32475 was determined by free fit, but
for   HIP~42910  the   outer  period   and  inclination   were  fixed.
Preliminary orbits  for this triple  with periods  of 106 and  9.06 yr
were   published  by   \citet{Horch2021};  they   disagree  with   all
measurements  available  at present.   A  preliminary  outer orbit  of
HIP~32475    with    $P=128.9$    yr     has    been    computed    by
\citet{Cvetkovic2020}.

The magnitude  difference between components  B and C of  HIP~42910 is
close to zero, so they can be swapped. An alternative to the eccentric
inner orbit with  $P=6.8$ yr could be a  highly inclined near-circular
orbit with  approximately double  period.  A quasi-circular  orbit was
fitted  to the  measurements of  B,C with  suitably changed  quadrants
($P=15.4$ yr, $a=0\farcs185$, $e=0.22$).   However, its agreement with
the measurements  is worse, and  the inner mass  sum of 1.28  \msun is
much  larger  than  allowed  by  the  absolute  magnitudes.   So,  the
eccentric orbit of  HIP~42910 B,C is the correct  choice. However, the
lack of measurements near its  periastron, when the subsystem is below
the SOAR resolution limit, does not fully constrain all elements.  For
this reason I fixed the inner  elements $\omega$ and $i$ to the values
that agree well with the data and  lead to the expected inner mass sum
of 0.73  \msun (the free fit  gives a slightly larger  mass sum).  The
next  inner  periastron   will  occur  in  2023.25,   and  the  latest
observation in 2023.0 confirms the decreasing separation.

\begin{deluxetable*}{l  l ccc   c  c  c}

\tabletypesize{\scriptsize}
%\tablenum{7}
\tablewidth{0pt}
\tablecaption{Photometry and Masses of HIP 32475 and 42910 
\label{tab:HDSptm}}
\tablehead{
%HIP & Param. & A+B+C & A-B & B-C & A      & B      & C \\
\colhead{HIP} &
\colhead{Parameter} &
\colhead{A+B+C} &
\colhead{A-B} &
\colhead{B-C} &
\colhead{A} &
\colhead{B} &
\colhead{C} 
}
\startdata
32475 & $I$ (mag)   & 6.95   & 3.71    & 0.32    & 7.02 & 10.59 & 10.91   \\
      & $M$ (\msun) & \ldots & \ldots  & \ldots  & 1.40 & 0.69  & 0.65 \\
42910 & $I$ (mag)   & 8.70   & 2.97    & 0.06   & 8.83  & 11.80 & 11.86 \\
      & $M$ (\msun) & \ldots & \ldots  & \ldots & 0.72  & 0.37  & 0.36  
\enddata
\end{deluxetable*}

Speckle  interferometry at  SOAR  gives reliable  measurements of  the
magnitude differences in  the spectral band close to  $I$.  These data
are assembled in Table~\ref{tab:HDSptm}.   For HIP~32475, the combined
$I$ magnitude should be  close to the $G$-band magnitude, 6.95 mag
(the color indices  are moderate). This assumption  and the isochrones
agree  with the  measured combined  $V$ and  $K$ magnitudes.   For the
redder star HIP~42910, I adopt the  combined $I=8.70$ mag based on the
following argument.   After splitting the flux  between components and
deriving  their absolute  $I$  magnitudes, I  use  the PARSEC  isochrone
\citep{PARSEC} for 1 Gyr and  solar metallicity to estimate the masses
and the combined $V$ and $K$  magnitudes of the system (10.10 and 7.10
mag,  respectively).   They  are  compared to  the  actually  measured
magnitudes (10.19  and 7.00  mag), and the  best agreement  is reached
with the adopted  combined $I$.

In  HIP~32475,  the  main  star  A,   of  F0IV  spectral  type,  is  a
$\gamma$~Dor   pulsating  variable   V830~Mon.   Its   photometrically
estimated mass, 1.40 \msun, is close  to the estimated mass sum of the
inner pair, 1.34 \msun, so both  inner and outer mass ratios are close
to one (a double twin); the inner orbit and parallax give the mass sum
of 1.44 \msun. The inner mass ratio is directly measured by the wobble
factor and, within errors, matches  the photometric masses.  The outer
mass ratio can  be checked by comparing the outer  orbital motion with
the proper motion anomaly, PMA \citep{Brandt2018}.  It equals $(-10.7,
-5.4)$ \masyr for the Gaia DR2  epoch of 2015.5, while the outer orbit
predicts an effective motion of  $(+20.6, +11.1)$ \masyr. The ratio of
PMA to  orbital motion  is close  to $-0.5$  and implies  $q_{\rm A,BC}
\approx 1$ if the photo-center motion  is attributed to star A and the
light of BC is neglected.

The estimated mass  of HIP~42910 A, 0.72  \msun, approximately matches
its spectral type  K7V.  This is also a double  twin.  The inner orbit
was tuned  to obtain  the expected  mass sum of  0.73 \msun,  as noted
above. The  poorly constrained outer orbit  yields a mass sum  of 1.45
\msun.

The similarity  of those two hierarchies  in terms of mass  ratios and
periods  contrasts  with very  different  character  of their  orbital
motions.  The orbits in HIP~32475 have moderate eccentricities and are
oriented  almost  face-on.   The  most  likely  value  of  the  mutual
inclination is  18\degr$\pm$6\degr (the  alternative value  of 60\degr
~would have  caused Lidov-Kozai  cycles that  would increase  the inner
eccentricity). The period ratio is small,  6.6, although it is not yet
accurately known.  Dynamical interactions in this low-hierarchy system
are expected  to be strong, and  a mean motion resonance  is possible.
This  triple  system  belongs  to  the  family  of  ``dancing  twins''
\citep{Dancingtwins}. On the other hand,  in HIP~42910 the inner orbit
has a large eccentricity of 0.95, and the mutual inclination is either
6\degr ~or 90\degr.

%---------------------------------------------------------
\section{Discussion}
\label{sec:disc}

\begin{figure}
\epsscale{1.1}
%\plotone{Scenario.eps}
\plotone{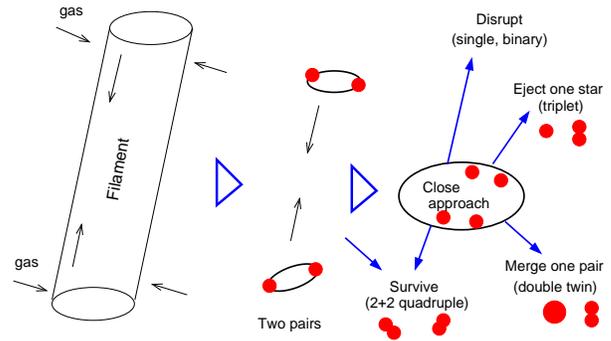}
\caption{Possible scenario of forming hierarchical systems with
  comparable-mass components: triplets, quadruplets, and double twins.  
\label{fig:scenario}
}
\end{figure}

It  is  firmly  established  that  masses of  stars  in  binaries  are
correlated instead of being chosen randomly \citep{DK13,Moe2017}. This
trend extends to hierarchical  systems. Quadruplets consisting of four
similar  stars arranged  in 2+2  hierarchy stand  apart as  a distinct
family of $\epsilon$~Lyr type, although their orbital separations span
a  wide  range \citep{Mult2021}.  Comparable  masses are  naturally
explained by gas accretion onto a binary that tends to equalize masses
and at the same time  shrinks the orbits \citep{TokMoe2020}. Extension
of this idea to triples helps to explain double twins where both inner
and outer mass ratios are close to one \citep{Dancingtwins}.  However,
existence of  hierarchical systems  of three similar  stars (triplets)
like $\epsilon$~Cha  with an  outer mass ratio  of 0.5  challenges the
accretion scenario.

A  possible path  to form  triplets is  via dynamical  decay of  a 2+2
quadruple    system.      This    scenario    is     illustrated    in
Figure~\ref{fig:scenario}.   The initial  condition is  a filament  of
dense  gas which  grows by  the  accretion flow  perpendicular to  its
axis. Inside  the filament, the flow  is directed along its  axis. Two
over-densities in  the filament form  two pairs of similar  stars with
orbits roughly  perpendicular to  the filament,  owing to  the angular
momentum of the incoming gas.  The total masses of both pairs are also
comparable  because they  formed  in the  same  filament and  experienced
comparable accretion rates.  The pairs  approach each other, driven by
mutual attraction and by  the center-of-mass velocities inherited from
the parental gas flow along the filament.

Close approach of  two pairs and their dynamical  interaction can lead
to  four different  outcomes  \citep{Antognini2016}.  In the  simplest
case, the decay products are just  single stars and binaries.  If only
one star is  ejected, a bound triple with three  similar components (a
triplet) could result.  Alternatively, one  pair can become very close
and merge, leaving a double twin.   Finally, a bound 2+2 quadruple can
emerge if  the dynamical interaction  was not  too violent or  did not
happen at all.  In all cases  the surviving hierarchies bear imprints of
chaotic  dynamics,   namely  eccentric   orbits  with   random  mutual
orientation.

The two massive triplets studied here ($\epsilon$~Cha and I~385) match
the proposed  scenario: their  inner orbits have  large eccentricities
and are not  aligned with the outer orbits.  HIP~42910,  a double twin
with eccentric inner  orbit, could be a merger  product.  In contrast,
the   architecture  of   the  double   twin  HIP~32475   with  aligned
quasi-circular orbits better matches  the accretion scenario discussed
in \citep{Dancingtwins}.

A  remarkable quadruple  system FIN~332  (WDS J18455+0530,  HIP 92027,
HR~7048,  the  ``tweedles'')  illustrates the  proposed  scenario.  It
consists  of four  nearly  equal A1V  type stars  in  a 2+2  hierarchy
\citep{FIN332}. Orbits of  the two inner twins have periods  of 28 and
40  yr  and large  eccentricities  (0.82  and 0.84);  moreover,  their
apsidal axes  point in approximately  same direction.  The  outer pair
($P  \sim 5$\,kyr)  moves  in  the opposite  sense  and  its orbit  is
definitely  misaligned   with  orbits   of  the  inner   pairs.   This
architecture strongly  suggests a past dynamical  interaction.  If one
of the  pairs in this  system were disrupted  and ejected a  star, the
result could resemble $\epsilon$~Cha or I~385.

If $\epsilon$~Cha is a product of a decaying 2+2 quadruple, one B-type
star should have been ejected. Assuming  an ejection speed of 30 \kms,
the star would have traveled a $\sim$150\,pc distance in 5\,Myr. It is
almost  hopeless to  search for  the ejected  star, it  can be  located
anywhere on the  sky. The phenomenon of runaway massive  stars is well
known, and  it is generally  accepted that many runaways  were ejected
from young unstable hierarchies \citep{Hoogerwerf2000}.  For effective
ejections, other  members of these  hierarchies must be  also massive,
and  this consideration  supports  the dynamical  scenario of  forming
massive triples and quadruplets.

However, the  scenario of triplet formation via decay of a 2+2
quadruple has a serious problem.  Ejection of one star with a velocity
$V$ causes  recoil of the  remaining triple  with a velocity  of $\sim
V/3$.  The facts that $\epsilon$~Cha is close to the neighboring stars
in the association and that I~385  is bound to another star C indicate
absence of  a fast recoil. Equal  masses in triplets can  be explained
alternatively  by   accretion  from  a  common   gas  reservoir  while
separations between the stars were still large and they moved randomly
through  the  parental  core without  mutual  dynamical  interactions;
otherwise, one star  would have been ejected without a  chance to grow
further, as discussed by \citet{Reipurth2000}. The N-body dynamics may
come into play later, when the  system have migrated to a more compact
configuration   and  the   gas  was   mostly  exhausted;   the  triple,
nevertheless, avoids  disruption and  continues to move  together with
its neighbors.

Study of relative motions in  hierarchical systems opens a fascinating
window  on  their  diversity   and  suggests  formation  via  several
channels, still  poorly explored.   Extension of such  work to  a much
larger  sample  of  hierarchies  is highly  desirable.  However,  long
periods  and  the  lack  of historic  measurements  severely  restrict
potential   samples.  Indirect   statistical  approaches   using  only
``instantaneous''     data    like     positions    and     velocities
\citep[e.g.][]{Hwang2022a} are  promising  for the dynamical study
of  typical  hierarchies with  separations  of  1--100 au.  Long-term
speckle monitoring of a large  number of resolved hierarchies combined
with precise Gaia astrometry will  provide input data for these future
investigations.

%---------------------------------------------------------
%\section{}
%\label{sec:}

%---------------------------------------------------------
%\section{}
%\label{sec:}

%\acknowledgements
\begin{acknowledgments}

The research  was funded by the  NSF's NOIRLab.  I thank  B.~Mason for
the analysis  of archival observations  of I~385.  This work  used the
SIMBAD   service  operated   by   Centre   des  Donn\'ees   Stellaires
(Strasbourg, France),  bibliographic references from  the Astrophysics
Data System  maintained by  SAO/NASA, and  the Washington  Double Star
Catalog maintained at  USNO.  This work has made use  of data from the
European     Space     Agency     (ESA)     mission     {\it     Gaia}
(\url{https://www.cosmos.esa.int/gaia}), processed  by the  {\it Gaia}
Data      Processing      and     Analysis      Consortium      (DPAC,
\url{https://www.cosmos.esa.int/web/gaia/dpac/consortium}).    Funding
for the DPAC has been provided by national institutions, in particular
the  institutions   participating  in  the  {\it   Gaia}  Multilateral
Agreement.  This  research has  made use  of the  services of  the ESO
Science Archive Facility.

\end{acknowledgments} 

\facility{SOAR, Gaia}

%---------------------------------------------------------
%\section{}
%\label{sec:}

%---------------------------------------------------------
%\section{}
%\label{sec:}

%---------------------------------------------------------
%\section{}
%\label{sec:}

%\subsection{}

%\bibliography{triples.bib}
%\bibliographystyle{aasjournal}

%\bibliographystyle{apj}
%\bibliography{chiron.bib}

\end{document}